\documentclass[prb,twocolumn,showpacs,floatfix,amsmath,amssymb,superscriptaddress]{revtex4}
\usepackage[pdftex]{graphicx}
\usepackage{latexsym}
\usepackage{graphicx}
\usepackage{times}
\usepackage{amsmath}
\usepackage{dcolumn}
\usepackage{subfigure}
\usepackage{latexsym,amsmath,amssymb,bm,euscript}
\usepackage{amsfonts}
\usepackage{hyperref}

\newcommand{\be}{\begin{equation}}
\newcommand{\ee}{\end{equation}}

\newcommand{\bs}{\begin{split}}
\newcommand{\es}{\end{split}}

\begin{document}

\title{Axion response in gapless systems}  
\author{Doron L. Bergman}
\affiliation{Physics Department, California Institute of Technology,
  MC 114-36, 1200 E. California Blvd., Pasadena, CA 91125}

\date{\today} 
 
\begin{abstract}
The strong topological insulator in 3D is expected to realize a quantized magneto-electric
response, the so-called axion response. However, many of the materials predicted to be 
topological insulators have turned out to be metallic, with bulk Fermi surfaces.
Following the result of Bergman et al. (Phys. Rev. B 82, 195417 (2010) ) that the helical surface
states of the topological insulator persist even when the band structure gap is closed,
we explore the fate of the magneto-electric response in such systems.
We find a non-quantized magneto-electric coupling remains once a bulk Fermi surface opens - a non-universal axion response.
More generally we find that higher dimensional analogs of the intrinsic anomalous Hall effect
appear for \emph{every} Chern form - non-quantized response coefficients for gapless systems, as opposed to 
quantized transport coefficients in gapped systems, both with a topological origin.
In particular, the non-quantized magneto-electric response in 3D descends
from the intrinsic anomalous Hall effect analog in 4D. 
\end{abstract} 
 
\maketitle 


Axion electrodynamics, long confined to the realm of high energy physics\cite{Wilczek:prl1987}, has been recognized recently as a possible effective phenomena in solids\cite{Qi:prb2008}. The strong topological insulator\cite{Fu:2007,Qi:prb2008,Essin:2009,TI_RMP}, at its boundary, is predicted to exhibit the physics of axion electrodynamics. The physical consequences of this fact are fantastic - ranging from a quantization of magneto electric coupling in materials with neither inversion symmetry breaking, nor time-reversal breaking, all the way to effective magnetic monopole image charges appearing when an external real charge is placed outside the material. 


Perhaps the greatest current issue with 3D topological insulator candidate materials, 
which has so far prevented putting to the test the interesting predictions mentioned 
above, is the fact that they have for the most part turned out to be 
metallic\cite{Hsieh:2009,Chen:2009,Hsieh:2010,Analytis:prb2010,Eto:prb2010}. 
The metallic states in these materials do show one characteristic feature of the topological 
insulator - the helical surface states. Theory tells us that topological states are only well 
defined when the spectrum of the bulk is gapped, yet in these materials the surface states 
coexist with a bulk Fermi surface.

%
%

Having noticed the coexistence of the helical surface states and bulk Fermi surfaces, and the
observed experimental fact that they seem at least mildly robust to disorder effects\cite{Hsieh:2009} 
Bergman et al.\cite{Bergman:prb2010} recently explored 
the physics of the surface states of a hybrid system of metallic and topological insulator bands, concluding that it is entirely possible to have helical surface states (a litmus test of the topological insulator) coexisting with a bulk Fermi surface.

\begin{figure}
	\centering			
		\subfigure[ Valleys in Graphene]{
		\includegraphics[width=1.5in]{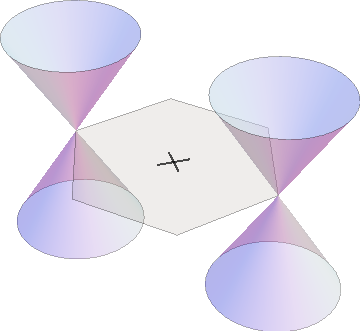}
			\label{fig:subfig1} }
		\subfigure[ Valleys in the helical metal]{
			\includegraphics[width=1.5in]{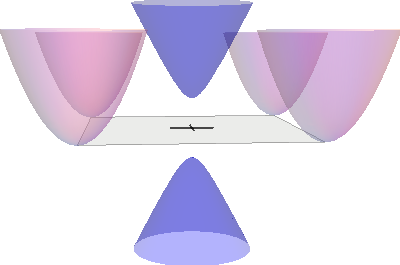}
		  \label{fig:subfig2} }				
\caption{Multi-valley low-energy description of Graphene and of the helical metal.
Panel \ref{fig:subfig2} shows the two Dirac cones of Graphene,
and panel \ref{fig:subfig1} shows a (dark-shaded) massive Dirac cone at the BZ center, and a (light-shaded) quadratic-dispersion Fermi pocket at the BZ corner. 
}
\label{fig:3band_numerics}
\end{figure}

The simplest way to think of these ``helical metals'' (bulk metals with helical surface states) 
is as follows. The minimal feature in the bulk band structure of a solid giving topological insulator behavior is a point in the first Brillouin zone around which the dispersion is approximately that of a massive Dirac point - dispersion
$E = \pm \sqrt{M^2 + {\bf q}^2}$. 
If a Fermi surface opens around some 
{\emph{other}} point in the Brillouin zone, an approximate low energy model for this band structure would consist of multiple valleys - separate valleys for the massive Dirac point, and for the Fermi surface(s) (see Fig.~\ref{fig:subfig2}). Much like the case of Graphene 
(see Fig.~\ref{fig:subfig1}), where the two Dirac points are to a good approximation independent because of the large momentum separation between them\cite{RMP_Graphene}, 
the massive Dirac point and the Fermi surface(s) will be to a good approximation independent
when they have a large separation in the Brillouin zone. From this simplistic picture we can understand that the helical surface states occur at the surface as a consequence of the massive Dirac point, and the Fermi surface simply does not always interfere with these surface states.
In particular, the surface states are the Jackiw-Rebbi bound states of the Dirac valley.
The band structure of Sb, expected to be a helical metal\cite{Bergman:prb2010} as it exhibits helical surface states,
can be described in low energies as a number of massive Dirac points, in addition to quadratic-dispersing 
Fermi surfaces - this would suggest that the simple multi-valley picture for the helical metal applies to Sb directly.

Using the simplistic multi-valley picture, where the Fermi surface and the massive Dirac point are for the most part independent, one would naively expect that the axion response of the 3D topological insulator would also appear in this (hybrid) system. In this article we explore this question in detail. We will find that the axion response can indeed appear, but its coefficient is no longer quantized.

A less obvious, yet even simpler model of the helical metal is the Dirac-like Hamiltonian
\be\label{BiSe_model}
H = -\mu + q_1 \gamma_1 + q_2 \gamma_2 + q_3 \gamma_3 + M \gamma_5
\; ,
\ee
where $\gamma_{1 \ldots 5}$ are the $4\times4$ $\gamma$-matrices,
and the chemical potential is set away from the gap in the spectrum $|\mu|>|M|$. 
In our convention the  $\gamma_{1 \ldots 4}$ are odd under both time reversal and inversion,
and $\gamma_5$ is even under both, so that this Hamiltonian is time-reversal and inversion symmetric.
This model qualitatively reproduces most features of the low-energy spectrum of 
Bi$_2$Se$_3$\cite{Analytis:prb2010} - namely a single massive Dirac point in the low-energy bulk spectrum, 
and a single massless Dirac surface Fermi surface. 
Both \eqref{BiSe_model} and minimal multi-valley models can be captured in a generic 4-band model
of the form $H = d_0({\bf q}) + d_a({\bf q}) \cdot \gamma_a$.
Using the 4-band model we will show axion response can exist in many gapless systems.

%
%


The axion response action 
$S_{\textrm{axion}} = \frac{e^2}{4\pi^2 \hbar c} \int_{{\bf x}, t} \phi {\vec E} \cdot {\vec B} = 
\frac{e^2}{8\pi^2 \hbar c} \int_{{\bf x}, t} \phi \epsilon_{\mu \nu \lambda \rho} \partial_{\mu} A_{\nu} \partial_{\lambda} A_{\rho}$,
(from this point on we will set $c \equiv \hbar \equiv e \equiv 1$)
requires the presence of an additional field $\phi$ perturbing the system.
For systems that are intrinsically time-reversal and inversion symmetric, the action must remain invariant under these symmetries.
The additional field $\phi$ therefore can either be odd under time-reversal and inversion, or if it is a \emph{compact} variable,
e.g. $\phi = \phi + 2\phi_0$ so that it is defined modulo $2 \phi_0$,
it may assume the values $\phi = 0$ or $\phi = \phi_0 = -\phi_0$.
Realizing a spatially varying non-compact $\phi$, that is odd under both time-reversal and inversion symmetries,
could be accomplished by magnetic impurities, which break both, and therefore engender a component that is odd under both.
With an impurity concentration varying on macroscopic length scales, we can realize a slowly varying $\phi$. 

The simplest way to see why we need the auxiliary field $\phi$ is as follows.
Naively, one could calculate a ${\vec E} \cdot {\vec B}$ response for any given material using linear response, 
which in any system with space-time translational invariance would yield
$S = \int_{x,x'} K_{\nu \rho}(x-x') A_{\nu}(x) A_{\rho}(x')
= \int_{x,x',q} K_{\nu \rho}(q) e^{i (x-x')\cdot q} A_{\nu}(x) A_{\rho}(x')
$,
where $x,x'$ denote space-time coordinates, and $q$ denotes a momentum-frequency vector. One needs 
to expand in small $q$ (a gradient expansion)
$K_{\nu \rho}(q) \approx \textrm{const} + (\textrm{linear in q}) 
+ \Omega_{\mu \nu \lambda \rho} q_{\mu} q_{\lambda}$
in order to recover the two derivatives in $\partial_{\mu} A_{\nu} \partial_{\lambda} A_{\rho}$. 
For the ${\vec E} \cdot {\vec B}$ term one would need $\Omega_{\mu \nu \lambda \rho} = \Omega_0 \epsilon_{\mu \nu \lambda \rho} + \ldots$.
Unfortunately, $q_{\mu} q_{\lambda}$ is symmetric in its two indices, and such a term would vanish. Therefore, linear response cannot give a 
nonvanishing ${\vec E} \cdot {\vec B}$ response. Remarkably, this also implies that even in a system with inversion and 
time reversal symmetries \emph{explicitly} broken, where the term ${\vec E} \cdot {\vec B}$
does not violate any symmetries, {\emph{no}} axion term can come from linear response in a translationally invariant system.
Consequently, we need the system to be inhomogeneous - and this is provided by the auxiliary field $\phi$, varying in space.
Response theory is easier to calculate assuming a translationally invariant system, so one can treat the $\phi$ field as an additional 
external field perturbing the system and calculate non-linear response instead of linear response.
In analogy with a gauge field, whose derivatives are the physical observable, only the derivatives of $\phi$ appear in the modified 
Maxwell's equations and have measurable physical consequences.


In the helical metal, the gapless version of the topological insulator, we will find that the coefficient in front of the 
axion action is modified to a non-universal value
\be\label{partial_axion}
S_{\textrm{axion}} = {\tilde c} \frac{1}{4\pi^2} \int_{{\bf x}, t} 
\phi {\vec E} \cdot {\vec B}
\; ,
\ee
with ${\tilde c} \neq 1$ and 
depending on the system details. 
The non-quantized magneto-electric response coefficient is analogous to the 
Hall conductivity in the intrinsic anomalous Hall effect (AHE) in 2D Ferromagnetic metals - 
because there is no gap, the Hall conductance is not quantized, even though it does have a topological 
origin - the Berry phase - and when opening up a bulk gap, the Hall conductance is quantized.


We will start by demonstrating that an analog of the AHE occurs in gapless non-interacting systems in \emph{any} odd spacetime dimension
$d=2n+1$, and from it, we will understand how the axion response appears in the helical metal. Being the first attempt at calculating this response, we
will ignore the effects of disorder and of finite temperature.

The Chern Simons forms in electrodynamic response in $d=2n+1$ space-time dimensions, 
are\cite{Golterman:1993}
\be\label{Chern_Form}
\begin{split}
S_{\textrm{Chern}} & =  
\chi_n \epsilon_{\mu_1 \ldots \mu_{2n+1}}
\int_x
A_{\mu_1} \partial_{\mu_2} A_{\mu_3} \ldots \partial_{\mu_{2n}} A_{\mu_{2n+1}}
\; .
\end{split}
\ee
The coefficient $\chi_n$ can be found using response theory
\be\label{Chern_coef}
\begin{split}
\chi_n =
\frac{(-1)^{n+1} \epsilon_{\mu_1 \ldots \mu_{2n+1}}  }{i (n+1)(2n+1)!} 
\int_{{\vec q}, \omega} Tr\left[ 
\prod_{j=1}^{2n+1} \left( G \partial_{\mu_j} G^{-1} \right) \right]
\; ,
\end{split}
\ee
where $G = \left[ i\omega_n - H({\bf q}) \right]^{-1}$ is the single particle Green's function,
and $\partial_{\mu} = \frac{\partial}{\partial q_{\mu}}$
with $q_0$ the real frequency, and $q_{1 \ldots 2n}$ being the Brillouin zone coordinates.

We will calculate this coefficient for a Hamiltonian of the form $H = d_0({\bf q}) + d_a({\bf q}) \cdot \gamma_a$ where 
the matrices $\gamma_{1 \ldots 2n+1}$ satisfy a Clifford algebra $ \left\{ \gamma_a, \gamma_b \right\} = 2 \delta_{a b}$,
as well as the identity
\be\label{Clifford2}
Tr\left[ \gamma_{a_1} \gamma_{a_2} \ldots \gamma_{a_{2n}} \gamma_{a_{2n+1}} \right] = 2n i \epsilon_{a_1 \ldots a_{2n+1}}
\; .
\ee
With $d_0 = 0$ this is precisely the form of a minimal model for the Chern insulator in $d=2+1$ and for the time-reversal invariant 4D Quantum Hall state in $d=4+1$\cite{Zhang:2001,Bernevig:2002} (which realizes the
strong topological insulator when reducing to $d=3+1$). With $d_0$ closing the gap in some regions of the Brillouin Zone, 
this describes the helical metal. First, we "normalize" the Hamiltonian by $R = |{\vec d}|$, by writing
$
H = R h = R \left( r_0 + r_a \gamma_a \right)
$,
where now ${\vec r}^2 = 1$
and
$
g = G R = R \left[ i \omega_n  - H \right]^{-1} = \left[ i \omega_n/R  - h \right]^{-1}
$.
It follows that
$
G \partial_{\mu} G^{-1} = g \partial_{\mu} g^{-1} + \partial_{\mu} \ln(R)
$,
in which the last term is a scalar, and so can be moved outside of the trace in \eqref{Chern_coef}.
As a result, when expanding the formula \eqref{Chern_coef}, any term where
more than one instance of $G \partial_{\mu} G^{-1} \rightarrow \partial_{\mu} \ln(R)$ appears, will
vanish, since $\partial_{\mu} \ln(R) \partial_{\mu'} \ln(R)$ is symmetric in the indices $\mu,\mu'$
and we have the completely anti-symmetric tensor multiplying the entire term.
From the remaining cases where zero or one instances of $G \partial_{\mu} G^{-1} \rightarrow \partial_{\mu} \ln(R)$ appear, we have
\be\label{normalized_exp}
\begin{split} &
\chi_n =  
\frac{(-1)^{n+1} \epsilon_{\mu_1 \ldots \mu_{2n+1}}  }{i (n+1)(2n+1)!} 
\int_q \Bigg\{ 
Tr\left[ \prod_{j=1}^{2n+1} \left( g \partial_{\mu_j} g^{-1} \right) \right]
\\  &
+ 
\sum_{j=1}^{2n+1}
 \partial_{\mu_j} \ln(R)
Tr\left[ \prod_{i \neq j} \left( g \partial_{\mu_i} g^{-1} \right) \right]
\Bigg\}
\; .
\end{split}
\ee
We will show the second term vanishes, by integrating by parts with respect to some
$q_{\mu_i}$. We first rewrite $ g \partial_{\mu_i} g^{-1} g = -  \partial_{\mu_i} g$
and then integrate by parts. Acting on the scalar term outside the trace, we get
$\partial_{\mu_i} \partial_{\mu_j} \ln(R)$, which is symmetric in the two indices
and thus vanishes at the hands of the anti-symmetric tensor.
Similarly, every instance in which the $\partial_{\mu_i}$ derivative acts on
any existing derivative term $\partial_{\mu_{\ell}} g^{-1} $ also yields an expression 
symmetric in its indices, and vanishes.
Otherwise, it must act on an instance of $g$, giving  
$\partial_{\mu_i} g = - g \partial_{\mu_i} g^{-1} g$
and producing pairs of expressions of the type
$
Tr\left[ 
\ldots g \partial_{\mu_i} g^{-1} g \partial_{\mu_{\ell}} g^{-1}  \ldots
\right]
+
Tr\left[ 
\ldots g \partial_{\mu_{\ell}} g^{-1} g \partial_{\mu_i} g^{-1}  \ldots
\right]
$
symmetric in $\mu_{i,\ell}$, and thus also vanishing.
There are $2n$ instances of $g$ in the trace, of which 2 get absorbed into 
the initial $\partial_{\mu_1} g$, leaving $2n-2$ instances of $g$
on which $\partial_{\mu_i}$ acts on after integrating by parts.
Since this is an even number, all such instances pair up as described 
above, and the whole sum vanishes.
We are left therefore, with the same expression as in \eqref{Chern_coef}
with $g$ replacing $G$. This will prove easier to calculate.
It is worth mentioning in passing that for the case of an insulator,
this is precisely the same procedure of deforming the bands to flat bands,
used in Ref.~\onlinecite{Qi:prb2008}.

Using the Clifford algebra, the Green's function can be re-written as
$ g = \frac{(i \omega/R - r_0) + r_a \gamma_a}{(i \omega/R - r_0)^2 - 1} $ (at $T=0$).
The terms $\partial_{\mu} g^{-1}$
will have a scalar part $\partial_{\mu} (i \omega/R - r_0)$,
which can leave the trace. If two or more instances of this scalar part appear, 
it will be symmetric in the two derivative indices, and vanish under the anti-symmetrization.
Therefore at most one such scalar part can appear, but this is also the minimal number, because one of the derivatives has to be with respect to frequency $\partial_0 g = 1/R $.
As a result, for all the other terms, we can ignore the scalar part of the current terms and take
$
g \partial_{\mu} g^{-1} = 
\frac{(i \omega/R - r_0) + r_a \gamma_a}{(i \omega/R - r_0)^2 - 1}
(-1) \gamma_b \partial_{\mu} r_b
$.
For every permutation of $\mu_{1 \ldots 2n+1}$ we can cyclically rotate the $\partial_0 g$ term to the left end in the trace formula, yielding
\be\label{Chern_coef_2}
\chi_n =
\frac{(-1)^{n+1} \epsilon_{\mu_1 \ldots \mu_{2n}}  }{i (n+1)(2n)!} 
\int_q Tr\left[ g \partial_0 g^{-1}
\prod_{j=1}^{2n} \left( g \partial_{\mu_j} g^{-1} \right) \right]
\; ,
\ee
where now the indices run only over momentum coordinates.
From ${\vec r}^2 = 1$ we get ${\vec r} \cdot \partial_{\mu} {\vec r} = 0$,
which we can use to find
$
g \partial_{\mu} g^{-1} g \partial_{\nu} g^{-1}= 
\frac{
\gamma_a \gamma_b \partial_{\mu} r_a \partial_{\nu} r_b
}{\left[ (i \omega/R - r_0)^2 - 1 \right]}
$.
Plugging this in \eqref{Chern_coef_2}, we arrive at
\be
\begin{split}
\chi_n =
\frac{(-1)^{n+1} \epsilon_{\mu_1 \ldots \mu_{2n}}  }{i (n+1)(2n)!} 
\int_q \frac{ Tr\left[ \gamma_{a_1} \ldots \gamma_{a_{2n}} g \right]
\prod_{j=1}^{2n}  \partial_{\mu_j} r_{a_j} 
}{\left[ (i \omega/R - r_0)^2 - 1 \right]^n R}
\; .
\end{split}
\ee
We first turn our attention to the trace over $\gamma$ matrices.
Due to the anti-symmetry in the $\mu_j$ indices, the product
of $\gamma_{a_{1 \ldots 2n}}$ should also be anti-symmetric in its indices.
In particular, those indices should be all distinct. 
It is then easy to show that the trace over these $2n$ matrices must vanish, 
$
Tr\left[ \gamma_{a_1} \ldots \gamma_{a_{2n}} \right] = 0
$.
Together with \eqref{Clifford2} we then find that
\be
\begin{split}
\chi_n & =
\frac{(-1)^{n+1} \epsilon_{\mu_1 \ldots \mu_{2n}} \epsilon_{a_1 \ldots a_{2n+1}} }{(n+1)(2n-1)!} 
\int_q 
\frac{
r_{a_{2n+1}} \times 
\prod_{j=1}^{2n}  \partial_{\mu_j} r_{a_j}
}{\left[ (i \omega/R - r_0)^2 - 1 \right]^{n+1} R}
\; .
\end{split}
\ee
Performing the $T=0$ Matsubara frequency integral, we get
$
\int_{-\infty}^{\infty} \frac{d\omega}{\left[ (i \omega - r_0)^2 - 1 \right]^{n+1}} = 
\frac{(-1)^{n+1} \sqrt{\pi} \Gamma(n+1/2)}{\Gamma(1+n)} \Theta(1-|r_0|)
$
where $\Gamma(n)$ is the Gamma-function, 
$\Theta$ is a step function, and we absorbed $R>0$ into $\omega$.
Using this, we find   
\be
\begin{split}
\chi_n & = 
\frac{\epsilon_{a_1 \ldots a_{2n+1}}}{4^n} 
{2n \choose n-1} 
\int_{\vec q}  \Theta(|{\vec d}| - |d_0|)  
r_{a_{2n+1}} \times \prod_{j=1}^{2n}  \partial_{q_j} r_{a_j}
\; ,
\end{split}
\ee
where $q_{1 \ldots 2n}$ run over the Brillouin zone coordinates. Only those areas of the Brillouin zone (BZ) where there is a local gap $|{\vec d}| > |d_0|$ contribute to the coefficient $\chi_n$. If there is a gap everywhere in the BZ,
the integral gives a quantized value. The quantization can be understood by realizing that the integrand calculates the (directed) infinitesimal hypersurface element, and the integral measures the total hypersurface swept by the 
mapping ${\vec r}({\bf q})$ from the BZ to the unit hypersphere $S_{2n}$ (the $2n$-sphere). This integral will give an integer 
times the hypersurface of the $2n$-sphere $\Omega_{2n} = \frac{2 (4\pi)^n n!}{(2n)!}$, giving us the transport coefficient
$
\chi_n = N \frac{2n}{(n+1)! (4\pi)^n}
$,
where $N$ is the topological integer.
For $n=1$ this describes the Hall conductivity, which in the gapped case gives the 
integer quantum Hall effect, and in the gapless case gives the intrinsic anomalous Hall effect contribution\cite{AHE_RMP}. 
For $n=2$ the gapped case gives the 4D quantum Hall effect\cite{Zhang:2001,Bernevig:2002},
while for the gapless case we get a non-universal transport coefficient.
The $n=2$ Chern response is the precursor of the axion response in 3D, and so our results strongly 
suggest that such a response can be found in 3D gapless systems, to which we turn now.

The simplest way to think of the axion response in 3 dimensions is to start with the the $n=2$ Chern form \eqref{Chern_Form},
and replace the \emph{extra} vector potential component $A_4$ with the auxiliary field $\phi$,
mentioned earlier, which couples to matter fields via the free-fermion propagator $G = G(q,\phi)$.
Note that since $A_4$ is odd under time reversal and inversion, $\phi$ must be odd as well, fitting its non-compact variety.
The auxiliary field $\phi$ couples to an extra current operator $J_4 = -\partial_{\phi} G^{-1}(q,\phi)|_{\phi = 0}$, and
all the other current operators are now $J_{\mu} = -\partial_{\mu} G^{-1}(q,\phi)|_{\phi = 0}$.
In addition, the 4th spatial dimension gets dropped, and the anti-symmetric symbols now include a 5th coordinate representing 
the extra field $\phi$. The derivation of the non-linear response follows through in a similar manner as described above
for the $d=2n+1$ case, using $G(q,\phi)$, except perhaps the integration by parts arguments used when converting to the ``normalized'' bands. However, the integration by parts used on \eqref{normalized_exp}, could be done with respect to any one of the $2n \geq 2$ different spacetime indices $\mu_i$, 
so we can easily choose one that is a frequency-momentum component rather than $\phi$.
The calculation therefore follows through \emph{exactly} in the same manner as in $d=4+1$,
and we get after a bit more algebra the coefficient from \eqref{partial_axion}
\be\label{main_result}
\begin{split}
{\tilde c} & = 
6 \pi^2
\epsilon_{a_1 \ldots a_5}
\int_{\vec q}  \Theta(|{\vec d}| - |d_0|)  
r_{a_1} 
\partial_{q_1} r_{a_2}
\partial_{q_2} r_{a_3}
\partial_{q_3} r_{a_4}
\partial_{\phi} r_{a_5}
\; .
\end{split}
\ee

\begin{figure}
	\centering
		\includegraphics[width=2.0in]{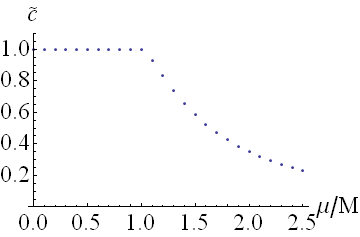}
	\caption{Calculation of the magneto electric response coefficient in a model system \eqref{toy_model} realizing both a gapped and gapless spectrum. 
	The transport coefficient ${\tilde c}$ is quantized as a long as a gap remains $|\mu|<M$. Once the gap is closed, ${\tilde c}$ is nonquantized.}
	\label{fig:simple_model_numbers_image}
\end{figure}

To demonstrate that ${\tilde c}$ can be nonzero in a gapless system, 
we will calculate it explicitly for \eqref{BiSe_model}, coupled to $\phi$ as follows
\be\label{toy_model}
H(\phi) = -\mu + q_1 \gamma_1 + q_2 \gamma_2 + q_3 \gamma_3 + \phi M \gamma_4 + M \gamma_5
\; .
\ee
Using the formula \eqref{main_result}, we can easily find
\be
{\tilde c} = \int_0^{\infty} \frac{3 M^2 q^2 dq}{(q^2 + M^2)^{5/2}} \Theta(q^2 + M^2 - \mu^2) 
\; .
\ee
Calculating numerically for a range of values for the chemical potential $\mu$ we find the results of Fig.~\ref{fig:simple_model_numbers_image},
which clearly indicate a quantized value ${\tilde c} = 1$ for the gapped case and a non-quantized value for the gapless case.

In summary, we have shown that in higher dimensions an analog of the 
intrinsic anomalous Hall effect exists in gapless systems closely related to 
Chern and topological insulators. As a consequence, the 3D descendant of the 
4D effect is a non-quantized axion response in 3D gapless systems.

The future directions following what we have shown here are interesting and numerous.
Borrowing from the what is known about the AHE\cite{AHE_RMP}, we expect that including disorder will bring about corrections to the axion response - both from a finite lifetime for quasiparticles, and from vertex corrections.
Given that the system is gapless, these corrections may be significant, and one must proceed with caution.
Still, given the robustness of the AHE in $d=2+1$ to disorder, it would be surprising if its higher dimensional analogs
(the non-quantized axion response among them) were completely washed out by disorder effects.

Another question that arises is how Coulomb screening will affect the electromagnetic system with this strange 
new response - contrary to the topological insulator, the helical metals will have free carriers in the bulk 
that can screen electric fields. We leave the resolution of these questions to future work.
The combined issues of disorder effects and screening make it difficult at this stage to predict how 
one would go about measuring experimentally the axion response in gapless systems.
However, probing AC response (where screening is no longer effective) may 
reveal the axion response.
We remain hopeful that the non-quantized axion response
could appear in helical metal materials doped with magnetic impurities,
perhaps in Bi$_{2-x}$Mn$_x$Te$_3$ \cite{Hor:prb2010}, doped to form a bulk Fermi surface.

DLB was supported by the Sherman Fairchild foundation,
and would like to acknowledge fruitful discussions with Gil Refael and Matthew P.A. Fisher.
During the completion of this work, the authors were made aware of the related work of 
Barkeshli and Qi\cite{Qi:2011}, discussing much, though not all, of the same physics.
Also, we thank G.E. Volovik, for pointing us to a discussion of topological effects 
in gapless systems\cite{Volovik:2005}.


\begin{thebibliography}{19}
\expandafter\ifx\csname natexlab\endcsname\relax\def\natexlab#1{#1}\fi
\expandafter\ifx\csname bibnamefont\endcsname\relax
  \def\bibnamefont#1{#1}\fi
\expandafter\ifx\csname bibfnamefont\endcsname\relax
  \def\bibfnamefont#1{#1}\fi
\expandafter\ifx\csname citenamefont\endcsname\relax
  \def\citenamefont#1{#1}\fi
\expandafter\ifx\csname url\endcsname\relax
  \def\url#1{\texttt{#1}}\fi
\expandafter\ifx\csname urlprefix\endcsname\relax\def\urlprefix{URL }\fi
\providecommand{\bibinfo}[2]{#2}
\providecommand{\eprint}[2][]{\url{#2}}

\bibitem[{\citenamefont{Wilczek}(1987)}]{Wilczek:prl1987}
\bibinfo{author}{\bibfnamefont{F.}~\bibnamefont{Wilczek}},
  \bibinfo{journal}{Phys. Rev. Lett.} \textbf{\bibinfo{volume}{58}},
  \bibinfo{pages}{1799} (\bibinfo{year}{1987}).

\bibitem[{\citenamefont{Qi et~al.}(2008)\citenamefont{Qi, Hughes, and
  Zhang}}]{Qi:prb2008}
\bibinfo{author}{\bibfnamefont{X.-L.} \bibnamefont{Qi}},
  \bibinfo{author}{\bibfnamefont{T.~L.} \bibnamefont{Hughes}},
  \bibnamefont{and} \bibinfo{author}{\bibfnamefont{S.-C.} \bibnamefont{Zhang}},
  \bibinfo{journal}{Phys. Rev. B} \textbf{\bibinfo{volume}{78}},
  \bibinfo{pages}{195424} (\bibinfo{year}{2008}).

\bibitem[{\citenamefont{Fu et~al.}(2007)\citenamefont{Fu, Kane, and
  Mele}}]{Fu:2007}
\bibinfo{author}{\bibfnamefont{L.}~\bibnamefont{Fu}},
  \bibinfo{author}{\bibfnamefont{C.~L.} \bibnamefont{Kane}}, \bibnamefont{and}
  \bibinfo{author}{\bibfnamefont{E.~J.} \bibnamefont{Mele}},
  \bibinfo{journal}{Phys. Rev. Lett.} \textbf{\bibinfo{volume}{98}},
  \bibinfo{pages}{106803} (\bibinfo{year}{2007}).

\bibitem[{\citenamefont{Essin et~al.}(2009)\citenamefont{Essin, Moore, and
  Vanderbilt}}]{Essin:2009}
\bibinfo{author}{\bibfnamefont{A.~M.} \bibnamefont{Essin}},
  \bibinfo{author}{\bibfnamefont{J.~E.} \bibnamefont{Moore}}, \bibnamefont{and}
  \bibinfo{author}{\bibfnamefont{D.}~\bibnamefont{Vanderbilt}},
  \bibinfo{journal}{Phys. Rev. Lett.} \textbf{\bibinfo{volume}{102}},
  \bibinfo{pages}{146805} (\bibinfo{year}{2009}).

\bibitem[{\citenamefont{Hasan and Kane}(2010)}]{TI_RMP}
\bibinfo{author}{\bibfnamefont{M.~Z.} \bibnamefont{Hasan}} \bibnamefont{and}
  \bibinfo{author}{\bibfnamefont{C.~L.} \bibnamefont{Kane}},
  \bibinfo{journal}{Rev. Mod. Phys.} \textbf{\bibinfo{volume}{82}},
  \bibinfo{pages}{3045} (\bibinfo{year}{2010}).

\bibitem[{\citenamefont{Hsieh et~al.}(2009)\citenamefont{Hsieh, Xia, Wray,
  Qian, Pal, Dil, Osterwalder, Meier, Bihlmayer, Kane et~al.}}]{Hsieh:2009}
\bibinfo{author}{\bibfnamefont{D.}~\bibnamefont{Hsieh}},
  \bibinfo{author}{\bibfnamefont{Y.}~\bibnamefont{Xia}},
  \bibinfo{author}{\bibfnamefont{L.}~\bibnamefont{Wray}},
  \bibinfo{author}{\bibfnamefont{D.}~\bibnamefont{Qian}},
  \bibinfo{author}{\bibfnamefont{A.}~\bibnamefont{Pal}},
  \bibinfo{author}{\bibfnamefont{J.~H.} \bibnamefont{Dil}},
  \bibinfo{author}{\bibfnamefont{J.}~\bibnamefont{Osterwalder}},
  \bibinfo{author}{\bibfnamefont{F.}~\bibnamefont{Meier}},
  \bibinfo{author}{\bibfnamefont{G.}~\bibnamefont{Bihlmayer}},
  \bibinfo{author}{\bibfnamefont{C.~L.} \bibnamefont{Kane}},
  \bibnamefont{et~al.}, \bibinfo{journal}{Science}
  \textbf{\bibinfo{volume}{323}}, \bibinfo{pages}{919} (\bibinfo{year}{2009}).

\bibitem[{\citenamefont{Chen et~al.}(2009)\citenamefont{Chen, Analytis, Chu,
  Liu, Mo, Qi, Zhang, Lu, Dai, Fang et~al.}}]{Chen:2009}
\bibinfo{author}{\bibfnamefont{Y.~L.} \bibnamefont{Chen}},
  \bibinfo{author}{\bibfnamefont{J.~G.} \bibnamefont{Analytis}},
  \bibinfo{author}{\bibfnamefont{J.-H.} \bibnamefont{Chu}},
  \bibinfo{author}{\bibfnamefont{Z.~K.} \bibnamefont{Liu}},
  \bibinfo{author}{\bibfnamefont{S.-K.} \bibnamefont{Mo}},
  \bibinfo{author}{\bibfnamefont{X.~L.} \bibnamefont{Qi}},
  \bibinfo{author}{\bibfnamefont{H.~J.} \bibnamefont{Zhang}},
  \bibinfo{author}{\bibfnamefont{D.~H.} \bibnamefont{Lu}},
  \bibinfo{author}{\bibfnamefont{X.}~\bibnamefont{Dai}},
  \bibinfo{author}{\bibfnamefont{Z.}~\bibnamefont{Fang}}, \bibnamefont{et~al.},
  \bibinfo{journal}{Science} \textbf{\bibinfo{volume}{325}},
  \bibinfo{pages}{178 } (\bibinfo{year}{2009}).

\bibitem[{\citenamefont{Hsieh et~al.}(2010)\citenamefont{Hsieh, Wray, Qian,
  Xia, Dil, Meier, Patthey, Osterwalder, Bihlmayer, Hor et~al.}}]{Hsieh:2010}
\bibinfo{author}{\bibfnamefont{D.}~\bibnamefont{Hsieh}},
  \bibinfo{author}{\bibfnamefont{L.}~\bibnamefont{Wray}},
  \bibinfo{author}{\bibfnamefont{D.}~\bibnamefont{Qian}},
  \bibinfo{author}{\bibfnamefont{Y.}~\bibnamefont{Xia}},
  \bibinfo{author}{\bibfnamefont{J.~H.} \bibnamefont{Dil}},
  \bibinfo{author}{\bibfnamefont{F.}~\bibnamefont{Meier}},
  \bibinfo{author}{\bibfnamefont{L.}~\bibnamefont{Patthey}},
  \bibinfo{author}{\bibfnamefont{J.}~\bibnamefont{Osterwalder}},
  \bibinfo{author}{\bibfnamefont{G.}~\bibnamefont{Bihlmayer}},
  \bibinfo{author}{\bibfnamefont{Y.~S.} \bibnamefont{Hor}},
  \bibnamefont{et~al.}, \bibinfo{journal}{New Journal of Physics}
  \textbf{\bibinfo{volume}{12}}, \bibinfo{pages}{125001}
  (\bibinfo{year}{2010}).

\bibitem[{\citenamefont{Analytis et~al.}(2010)\citenamefont{Analytis, Chu,
  Chen, Corredor, McDonald, Shen, and Fisher}}]{Analytis:prb2010}
\bibinfo{author}{\bibfnamefont{J.~G.} \bibnamefont{Analytis}},
  \bibinfo{author}{\bibfnamefont{J.-H.} \bibnamefont{Chu}},
  \bibinfo{author}{\bibfnamefont{Y.}~\bibnamefont{Chen}},
  \bibinfo{author}{\bibfnamefont{F.}~\bibnamefont{Corredor}},
  \bibinfo{author}{\bibfnamefont{R.~D.} \bibnamefont{McDonald}},
  \bibinfo{author}{\bibfnamefont{Z.~X.} \bibnamefont{Shen}}, \bibnamefont{and}
  \bibinfo{author}{\bibfnamefont{I.~R.} \bibnamefont{Fisher}},
  \bibinfo{journal}{Phys. Rev. B} \textbf{\bibinfo{volume}{81}},
  \bibinfo{pages}{205407} (\bibinfo{year}{2010}).

\bibitem[{\citenamefont{Eto et~al.}(2010)\citenamefont{Eto, Ren, Taskin,
  Segawa, and Ando}}]{Eto:prb2010}
\bibinfo{author}{\bibfnamefont{K.}~\bibnamefont{Eto}},
  \bibinfo{author}{\bibfnamefont{Z.}~\bibnamefont{Ren}},
  \bibinfo{author}{\bibfnamefont{A.~A.} \bibnamefont{Taskin}},
  \bibinfo{author}{\bibfnamefont{K.}~\bibnamefont{Segawa}}, \bibnamefont{and}
  \bibinfo{author}{\bibfnamefont{Y.}~\bibnamefont{Ando}},
  \bibinfo{journal}{Phys. Rev. B} \textbf{\bibinfo{volume}{81}},
  \bibinfo{pages}{195309} (\bibinfo{year}{2010}).

\bibitem[{\citenamefont{Bergman and Refael}(2010)}]{Bergman:prb2010}
\bibinfo{author}{\bibfnamefont{D.~L.} \bibnamefont{Bergman}} \bibnamefont{and}
  \bibinfo{author}{\bibfnamefont{G.}~\bibnamefont{Refael}},
  \bibinfo{journal}{Phys. Rev. B} \textbf{\bibinfo{volume}{82}},
  \bibinfo{pages}{195417} (\bibinfo{year}{2010}).

\bibitem[{\citenamefont{Castro~Neto et~al.}(2009)\citenamefont{Castro~Neto,
  Guinea, Peres, Novoselov, and Geim}}]{RMP_Graphene}
\bibinfo{author}{\bibfnamefont{A.~H.} \bibnamefont{Castro~Neto}},
  \bibinfo{author}{\bibfnamefont{F.}~\bibnamefont{Guinea}},
  \bibinfo{author}{\bibfnamefont{N.~M.~R.} \bibnamefont{Peres}},
  \bibinfo{author}{\bibfnamefont{K.~S.} \bibnamefont{Novoselov}},
  \bibnamefont{and} \bibinfo{author}{\bibfnamefont{A.~K.} \bibnamefont{Geim}},
  \bibinfo{journal}{Rev. Mod. Phys.} \textbf{\bibinfo{volume}{81}},
  \bibinfo{pages}{109} (\bibinfo{year}{2009}).

\bibitem[{\citenamefont{Golterman et~al.}(1993)\citenamefont{Golterman, Jansen,
  and Kaplan}}]{Golterman:1993}
\bibinfo{author}{\bibfnamefont{M.~F.~L.} \bibnamefont{Golterman}},
  \bibinfo{author}{\bibfnamefont{K.}~\bibnamefont{Jansen}}, \bibnamefont{and}
  \bibinfo{author}{\bibfnamefont{D.~B.} \bibnamefont{Kaplan}},
  \bibinfo{journal}{Physics Letters B} \textbf{\bibinfo{volume}{301}},
  \bibinfo{pages}{219 } (\bibinfo{year}{1993}), ISSN \bibinfo{issn}{0370-2693}.

\bibitem[{\citenamefont{Zhang and Hu}(2001)}]{Zhang:2001}
\bibinfo{author}{\bibfnamefont{S.-C.} \bibnamefont{Zhang}} \bibnamefont{and}
  \bibinfo{author}{\bibfnamefont{J.}~\bibnamefont{Hu}},
  \bibinfo{journal}{Science} \textbf{\bibinfo{volume}{294}},
  \bibinfo{pages}{823} (\bibinfo{year}{2001}).

\bibitem[{\citenamefont{Bernevig et~al.}(2002)\citenamefont{Bernevig, Chern,
  Hu, Toumbas, and Zhang}}]{Bernevig:2002}
\bibinfo{author}{\bibfnamefont{B.~A.} \bibnamefont{Bernevig}},
  \bibinfo{author}{\bibfnamefont{C.-H.} \bibnamefont{Chern}},
  \bibinfo{author}{\bibfnamefont{J.-P.} \bibnamefont{Hu}},
  \bibinfo{author}{\bibfnamefont{N.}~\bibnamefont{Toumbas}}, \bibnamefont{and}
  \bibinfo{author}{\bibfnamefont{S.-C.} \bibnamefont{Zhang}},
  \bibinfo{journal}{Annals of Physics} \textbf{\bibinfo{volume}{300}},
  \bibinfo{pages}{185 } (\bibinfo{year}{2002}), ISSN \bibinfo{issn}{0003-4916}.

\bibitem[{\citenamefont{Nagaosa et~al.}(2010)\citenamefont{Nagaosa, Sinova,
  Onoda, MacDonald, and Ong}}]{AHE_RMP}
\bibinfo{author}{\bibfnamefont{N.}~\bibnamefont{Nagaosa}},
  \bibinfo{author}{\bibfnamefont{J.}~\bibnamefont{Sinova}},
  \bibinfo{author}{\bibfnamefont{S.}~\bibnamefont{Onoda}},
  \bibinfo{author}{\bibfnamefont{A.~H.} \bibnamefont{MacDonald}},
  \bibnamefont{and} \bibinfo{author}{\bibfnamefont{N.~P.} \bibnamefont{Ong}},
  \bibinfo{journal}{Rev. Mod. Phys.} \textbf{\bibinfo{volume}{82}},
  \bibinfo{pages}{1539} (\bibinfo{year}{2010}).

\bibitem[{\citenamefont{Hor et~al.}(2010)\citenamefont{Hor, Roushan,
  Beidenkopf, Seo, Qu, Checkelsky, Wray, Hsieh, Xia, Xu et~al.}}]{Hor:prb2010}
\bibinfo{author}{\bibfnamefont{Y.~S.} \bibnamefont{Hor}},
  \bibinfo{author}{\bibfnamefont{P.}~\bibnamefont{Roushan}},
  \bibinfo{author}{\bibfnamefont{H.}~\bibnamefont{Beidenkopf}},
  \bibinfo{author}{\bibfnamefont{J.}~\bibnamefont{Seo}},
  \bibinfo{author}{\bibfnamefont{D.}~\bibnamefont{Qu}},
  \bibinfo{author}{\bibfnamefont{J.~G.} \bibnamefont{Checkelsky}},
  \bibinfo{author}{\bibfnamefont{L.~A.} \bibnamefont{Wray}},
  \bibinfo{author}{\bibfnamefont{D.}~\bibnamefont{Hsieh}},
  \bibinfo{author}{\bibfnamefont{Y.}~\bibnamefont{Xia}},
  \bibinfo{author}{\bibfnamefont{S.-Y.} \bibnamefont{Xu}},
  \bibnamefont{et~al.}, \bibinfo{journal}{Phys. Rev. B}
  \textbf{\bibinfo{volume}{81}}, \bibinfo{pages}{195203}
  (\bibinfo{year}{2010}).

\bibitem[{\citenamefont{Barkeshli and Qi}(2011)}]{Qi:2011}
\bibinfo{author}{\bibfnamefont{M.}~\bibnamefont{Barkeshli}} \bibnamefont{and}
  \bibinfo{author}{\bibfnamefont{X.-L.} \bibnamefont{Qi}}
  (\bibinfo{year}{2011}), \bibinfo{note}{arxiv/1101.3104}.

\bibitem[{\citenamefont{Klinkhamer and Volovik}(2005)}]{Volovik:2005}
\bibinfo{author}{\bibfnamefont{F.}~\bibnamefont{Klinkhamer}} \bibnamefont{and}
  \bibinfo{author}{\bibfnamefont{G.}~\bibnamefont{Volovik}},
  \bibinfo{journal}{Int. J. of Mod. Phys. A} \textbf{\bibinfo{volume}{20}},
  \bibinfo{pages}{2795} (\bibinfo{year}{2005}).

\end{thebibliography}

\end{document}